\documentclass{article}
\usepackage{spconf,amsmath,graphicx}
\usepackage{amssymb}
\usepackage{url}
\usepackage{booktabs}
\usepackage[linesnumbered,ruled]{algorithm2e}
\usepackage{comment}
\usepackage{color}
\usepackage{etoolbox}
\usepackage{flushend}
\newbool{inccomment}
\setbool{inccomment}{true}
\usepackage{threeparttable}

\newcommand{\hl}[1]{\ifbool{inccomment}{{\color{blue}#1}}{}}
\newcommand{\JZ}[1]{\ifbool{inccomment}{{\color{green}#1}}{}}
\newcommand{\JH}[1]{\ifbool{inccomment}{{\color{yellow}#1}}{}}
\newcommand{\MD}[1]{\ifbool{inccomment}{{\color{red}#1}}{}}
\usepackage[belowskip=-3.2pt,aboveskip=0pt]{caption}

\newcommand{\dave}[1]{\ifbool{}{{\color{blue}#1}}{}}

\title{Structural sparsification for Far-field Speaker Recognition with Intel\textregistered\space GNA}
\name{Jingchi Zhang$^{\star}$ \quad Jonathan Huang$^{\dagger}$ \quad Michael Deisher$^{\diamond}$ \quad Hai Li$^{\star}$ \quad Yiran Chen$^{\star}$}
\address{$^{\star}$ Duke University, Durham, North Carolina, USA \\
$^{\dagger}$Intel Corporation, Santa Clara, California USA \\
$^{\diamond}$Intel Corporation, Hillsboro, Oregon, USA}
\begin{document}
%
\maketitle
\begin{abstract}
Recently, deep neural networks (DNN) have been widely used in speaker recognition area. In order to achieve fast response time and high accuracy, the requirements for hardware resources increase rapidly. However, as the speaker recognition application is often implemented on mobile devices, it is necessary to maintain a low computational cost while keeping high accuracy in far-field condition. In this paper, we apply structural sparsification on time-delay neural networks (TDNN) to remove redundant structures and accelerate the execution. On our targeted hardware, our model can remove 60\% of parameters and only slightly increasing equal error rate (EER) by 0.18\% while our structural sparse model can achieve more than $1.5\times$ speedup.
\end{abstract}

\begin{keywords}
 structural sparsification, far-field speaker recognition, time-delay neural network
\end{keywords}


\section{Introduction}
\label{sec:intro}
Far-field speaker recognition has gained much interest in the research community, with its prevalent applications in consumer devices such smart speakers and smartphones.  The far-field condition presents additional challenges in speaker recognition, due to the severity of reverberation and background noise.  Similar to automatic speech recognition, deep learning acoustic features have shown great improvements in these conditions compared to prior techniques.  A number of speaker recognition systems based on deep neural network (DNN) embeddings have been reported in the literature~\cite{heigold2016end}\cite{li2017deep}\cite{xvector}. More recently, SRI developed the VOiCES dataset~\cite{richey2018voices} specifically for far-field speaker recognition, and showed their DNN embeddings significantly outperformed the i-vector systems~\cite{nandwana2018robust}.

The objective of our work is to develop a speaker recognition system robust to the far-field channel conditions, using advanced model training methodology.  Furthermore, we designed the system to be simple to perform inference using ultra-low power accelerators such as the Intel\textregistered\space GNA~\cite{gna}. In contrast to the popular approach of using probabilistic linear discriminant analysis (PLDA)~\cite{ioffe2006probabilistic} in the back-end, our system only relies on the simple cosine distance for scoring. This allows for the computations to be performed end-to-end on the accelerator. Finally, to get the best model size efficiency, the crux of the paper will focus on the application of structural sparsification to our DNN model. Applying suitable sparse granularity on the model could reduce the latency of the model inference and improve the performance of the real-time speech recognition.

There have been extensive studies on accelerating DNN models. Pruning~\cite{Han} and sparsity methods~\cite{Liu} can effectively reduce the size of CNN models while keeping the performance similar to the original models. However, randomly distributed zeros in models do not have benefit for execution on hardware.~\cite{zhang2019learning} elaborates the benefit of structural sparsity over non-structural sparsity on locality and parallelism during hardware execution. To force zero parameters to form a regular arrangement, structural sparsity~\cite{Wei_cnn} is proposed for CNNs to learn sparse structures like channel and filter.

In order to reduce the model size and the inference time, we apply a structural sparsity learning method to speaker recognition models. The sparse structure we achieve is computationally friendly to specific hardware. Specifically, we add a group Lasso~\cite{yuan2006model} penalty to the loss function, where the group is the structure desired to be sparse. The sparse model performance is the same or even better compared to the baseline with fewer non-zero parameters. Also, we test our method on three different sparse granularity levels and found that under the same number of non-zero parameters, models with smaller granularity achieve lower equal error rate (EER) than models with larger granularity. Sparse model performance exceeds that of dense models regardless of the granularity with the same number of non-zero parameters.

\section{Related Work}
\label{sec:relate}
Computational acceleration methods have been heavily explored for the past years. Pruning and sparsification have proven effective at removing redundant parameters and structures. In~\cite{Han}\cite{han2015deep}, pruning connections of fully connected layers was proved effective at reducing the size of \textit{Alexnet} and \textit{VGG-16}. However, most of the computation and parameters are from convolution layers. From this perspective, Wei \textit{et al.}~\cite{Wei_cnn} propose a framework that can reduce model size by eliminating redundant structures in CNNs such as filters or channels. They claimed to achieve $3.1\times$ speedup on \textit{Alexnet} on GPU while keep the accuracy the same.

For speech recognition tasks, recurrent neural network (RNN) and long short-term memory (LSTM) models are widely used. It is more difficult to learn sparse structures for these models because the structures usually contain information on time sequences. Eliminating those structures would have more impact on performance. Narang \textit{et al.}~\cite{Narang} conducted Connection Pruning for RNNs and reduced 90\% of connections. Wei \textit{et al.}~\cite{intrinsic} further applied group Lasso regularization on LSTMs and achieved $10.59\times$ speedup without perplexity loss. Zhang \textit{et al.}~\cite{zhang2019learning} also extended the structural sparsity learning method to LSTM models for speech recognition and removed 72.5\% parameters with negligible accuracy loss.  

\section{Methodology}
\label{sec:methid}
\subsection{Model topology}
\label{ssec:topology}
This work is based the x-vector model structure~\cite{xvector}, with some simplifications. Compared to the original x-vector model, our architecture, shown on Table \ref{table:topology}, has increased the input feature dimension from 24 to 40, reduced the pooling dimension from 1500 to 512, removed a fully-connected layer between the embedding and speaker output layers, and reduced the embedding dimension from 512 to 256. In our testing, these modifications did not degrade recognition performance and had much lower complexity. We use this topology as the baseline for structural sparsity learning. Also, in this particular case, TDNN can be written as a one-dimension convolution, so we implemented the model as a 5-layer CNN. 

The softmax output is only used for model training purposes; for speaker enrollment and verification, the DNN embedding is taken at the output of Segment6 on Table \ref{table:topology}.  One speaker embedding is computed for an entire utterance, regardless of length. We use cosine distance of this 256-dimension embedding vectors between enrollment and test utterances to produce the speaker recognition score.  

\begin{table}[tb]
\caption{Model configuration}
\label{table:topology}
\small
\centering
\setlength{\tabcolsep}{2.5pt}
\vspace{3pt}
\begin{tabular}{|c|c|c|c|}
\toprule
 & layer context & Affine & Convolution \\
\hline
Layer1 & [t-2,t+2] & 200$\times$512 & 512 40$\times$5 \\
\hline
Layer2 & \{t-2,t,t+2\} & 1536$\times$512 & 512 512$\times$3 \\
\hline
Layer3 & \{t-2,t,t+2\} & 1536$\times$512 & 512 512$\times$3 \\
\hline
Layer4 & \{t\} & 512$\times$512 & 512 512$\times$1 \\
\hline
Layer5 & \{t\} & 512$\times$512 & 512 512$\times$1 \\
\hline
Stats pooling & [0,T) & 512T$\times$1024 &N/A  \\
\hline
Segment6 & \{0\} & 1024$\times$256 & N/A \\
\hline
Softmax & \{0\} & 256$\times$N & N/A \\
\bottomrule
\end{tabular}
\begin{tablenotes}
       \footnotesize
       \item[1] N denotes the number of training speakers.
     \end{tablenotes}
\end{table}

\subsection{Loss function}
\label{ssec:loss}
While the conventional softmax loss works reasonably well for training speaker embeddings, it is specifically designed for classification, not verification tasks. Speaker recognition systems trained with softmax loss typically use PLDA in the backend to improve separation between speakers. The triplet loss function, which is designed to reduce intra-speaker and increase inter-speaker distance, has shown to be more effective for speaker recognition \cite{li2017deep}. Likewise, the end-to-end loss \cite{heigold2016end} has better performance than softmax. The downside to these kinds of losses is that the training infrastructure is significantly more complicated than one used for supervised learning with softmax. In a prior study ~\cite{Huang2019IntelFFSID}, we explored the use of several recently proposed loss functions that were first introduced in face recognition research. These loss functions are drop-in replacements for softmax, thus modification to training code is simple with little overhead in training speed. We found Additive Margin Softmax (AM-softmax) \cite{wang2018additive} to perform best in the far-field test set, and incorporating PLDA did not improve performance against the simpler cosine distance. The elimination of the PLDA in the inference pipeline makes the entire model easy to deploy to target hardware, with the help of tools such as the Intel\textregistered\space Distribution of OpenVINO\textsuperscript{TM} toolkit~\cite{openvino}. 

\subsection{Training details}
\label{ssec:pipeline}
We describe our training pipeline as a three step process:
\begin{enumerate}
  \item \textbf{Baseline model training:} We find that we get significantly better results when we start the sparsification process with a well-trained dense model. We train the model with AM-softmax loss, SGD optimizer learning rate decaying from 0.01 to 0.0001 in 30 epochs with cosine annealing.  The weight decay and batch size are set to 1e-6 and 256, respectively.  For each batch, we select random segments of training utterances between 2.5 to 3.0 seconds.  These settings, except for the number of epochs, are used in subsequent steps.  The output of this step is the best dense model we can produce, and it also serves as a baseline to measure EER against.
  \item \textbf{Learning sparse structure:} We use the model from step 1 to initialize the dense model, and trained 20 epochs with the group Lasso regularization together with the AM-softmax loss:
  \begin{equation}
\label{equ:gl}
E(w)=E_D(w)+\lambda\cdot\sum_{k=1}^{K}||w_k||_2,
\end{equation}
where the first term $E_D(w)$ is the original AM-softmax loss function, and the second term is the contribution from the group Lasso loss function.  The group Lasso loss is essentially the summation of $K$ (the total number of groups) L2 norm of group weights $w_k$ in predefined groups (e.g. chunks of 8 or 16, or entire convolution filter).  It rewards to total loss function for forcing low values to group weights.  The coefficient $\lambda$ controls the balance between AM-softmax loss and group Lasso loss. This step produces sparse structures by training on the new loss function $E(w)$. Groups with L2 values below a threshold are set to 0, and discarded in the learning process for the next step.
  \item \textbf{Fine-tuning:} Lastly, we fine-tune the training for 20 epochs on the sparse model produced by step 2 using only AM-softmax loss. 
\end{enumerate}
More detail on step 2 and step 3 can be found in~\cite{zhang2019learning}.

\section{Experimental setup}
\subsection{Datasets and augmentation}
\label{ssec:data}
We use VoxCeleb 1 and 2 ~\cite{nagrani2017voxceleb}~\cite{chung2018voxceleb2} to train the system.  These datasets have 7323 identities combined.   We perform 9x data augmentation plus original clean speech to produce 12.7 million training utterances.  For each data augmentation, we randomly choose from 2000 room impulse responses generated from Pyroomacoustics ~\cite{scheibler2018pyroomacoustics}, and add randomly selected background noise from MUSAN ~\cite{Snyder15-MAM} and AudioSet ~\cite{gemmeke2017audio}.  For the test set, we used the VOiCES far-field dataset~\cite{richey2018voices}, which we believe captures the essence of challenging channel conditions. For all speech utterances, we use 40-dimension log-mel filterbanks, with 3-second sliding window mean subtraction.

\subsection{Hardware implementation}
\label{ssec:hardware}

This work is targeting TDNN inference on the Intel\textregistered\space Gaussian \& Neural Accelerator (GNA)~\cite{GNA2}. Intel\textregistered\space GNA is designed for continuous inference with neural networks on edge devices with high performance and very low power consumption. Since Intel\textregistered\space GNA fetches weight matrices in 16-byte chunks of \texttt{int8} or \texttt{int16} weights, we investigated structural sparsity on chunks of 8 \texttt{int16} elements or 16 \texttt{int8} elements.  Inference measurements were made on an Intel\textregistered\space Celeron\textregistered\space Processor J4005 with Intel\textregistered\space GNA inside.




\section{Results}
\label{sec:results}

In the experiments, the sparsity of filters is defined as the number of zero filters over all filters, while the sparsity of chunks is defined as the the number of zero chunks over all chunks. We applied the sparsity learning only to layers 1-4.  Our experiments showed that layers 5 and above were reluctant to achieve sparsity.  We suspect that this is because near the output of the network, the hidden representations contain high density of information for speaker recognition.  This seems to happen at the input of the stats pooling layer.

\subsection{Result analysis}
\label{ssec:analysis}

The experimental results are shown in Table~\ref{tab:filters}. We applied the structural sparsity on filters and chunks. Filter sparsity can be deployed on all hardware without any special modification. While applying sparsity on chunk-8 and chunk-16 are targeted at Intel\textregistered\space GNA. Also, we run experiments on dense models to compare the performance of sparse models and dense models.

\begin{figure}[tb]
\begin{minipage}[b]{0.3\linewidth}
  \centering
  \centerline{\includegraphics[width=3cm]{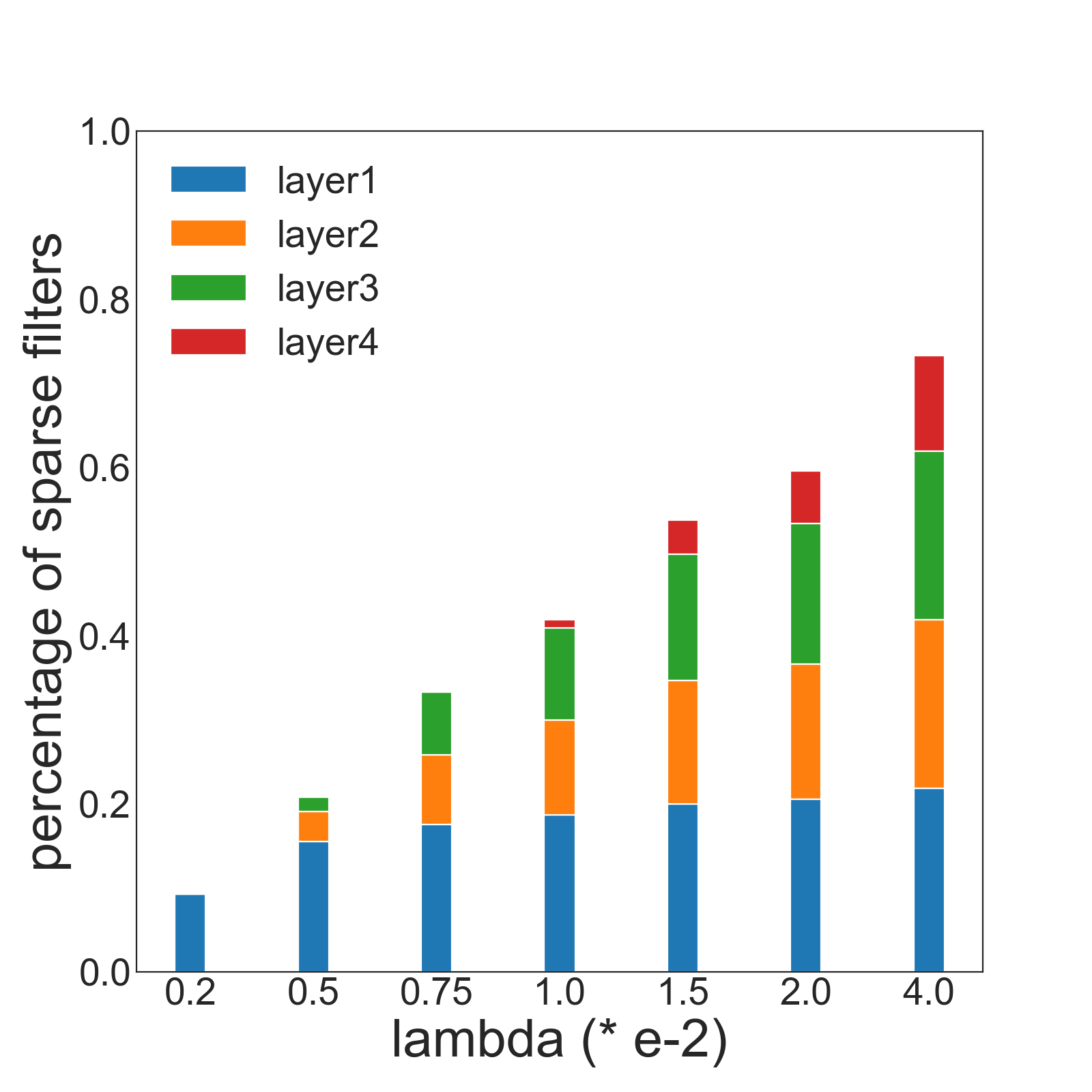}}
  \centerline{(a) \textit{filter}}
\end{minipage}
\hfill
\begin{minipage}[b]{0.3\linewidth}
  \centering
  \centerline{\includegraphics[width=3cm]{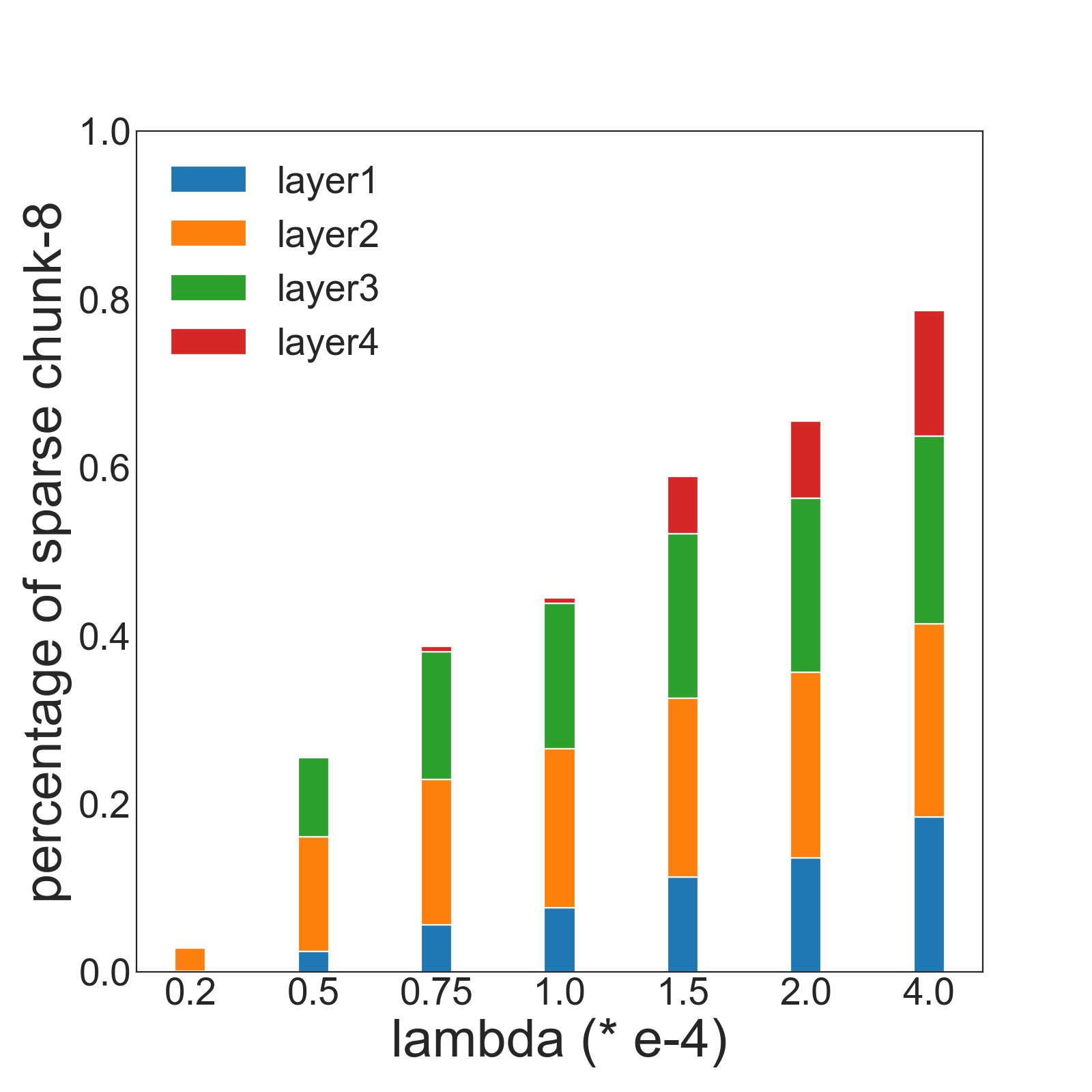}}
  \centerline{(b) \textit{group-8}}
\end{minipage}
\hfill
\begin{minipage}[b]{0.3\linewidth}
  \centering
  \centerline{\includegraphics[width=3cm]{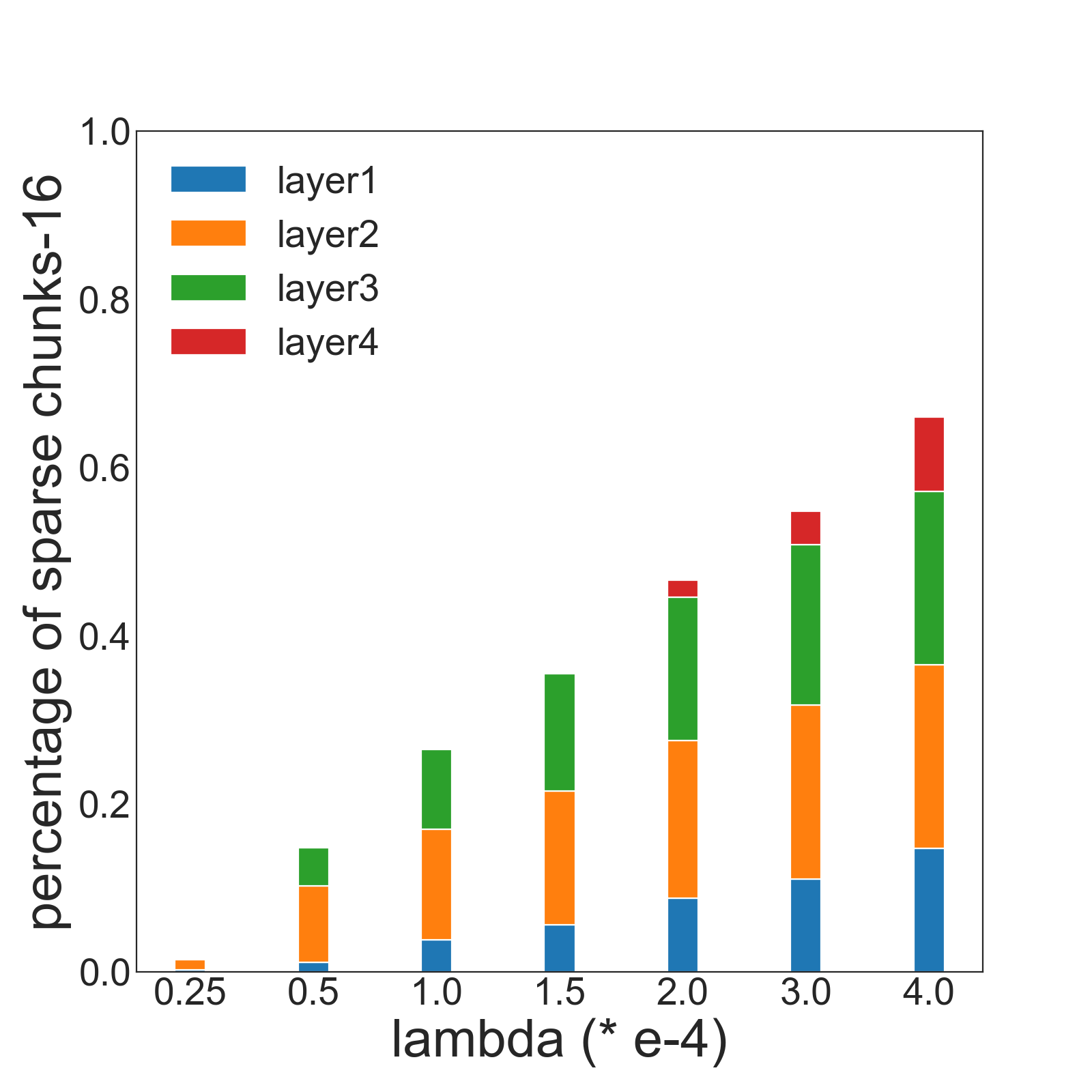}}
  \centerline{(b) \textit{group-16}}
\end{minipage}
\caption{Relationship between $\lambda$ and structural sparsity}
\label{fig:lambda_sparsity}
\end{figure}

Figure~\ref{fig:lambda_sparsity} is the visualization of the relationship between the coefficient $\lambda$ and sparsity in each layer. Y-axis denotes the overall percentage of sparsity in four layers. It is shown clearly when $\lambda$ increases, the sparsity increases. However, the sparsity growth in each layer is different. Filter sparsity shows a different growth trend from chunk sparsity. In Figure~\ref{fig:lambda_sparsity}(a), sparse filters in the first layer (blue bar) account for much of the overall sparsity. However, in Figure~\ref{fig:lambda_sparsity}(b) and (c), the first layer is not very sparse while layers 2 and 3 have a majority of chunks learned to be zero. We suspect that the low sparsity in layer 1 is due to the denser spectral input dimension compared to other layers; and that in layer 4 the output representation is becoming more relevant for the speaker recognition task, thus having making the network sparse here would result in higher penalty on the AM-softmax loss. 

\begin{figure}[tb]
\begin{minipage}[b]{0.45\linewidth}
  \centering
  \centerline{\includegraphics[width=4.3cm]{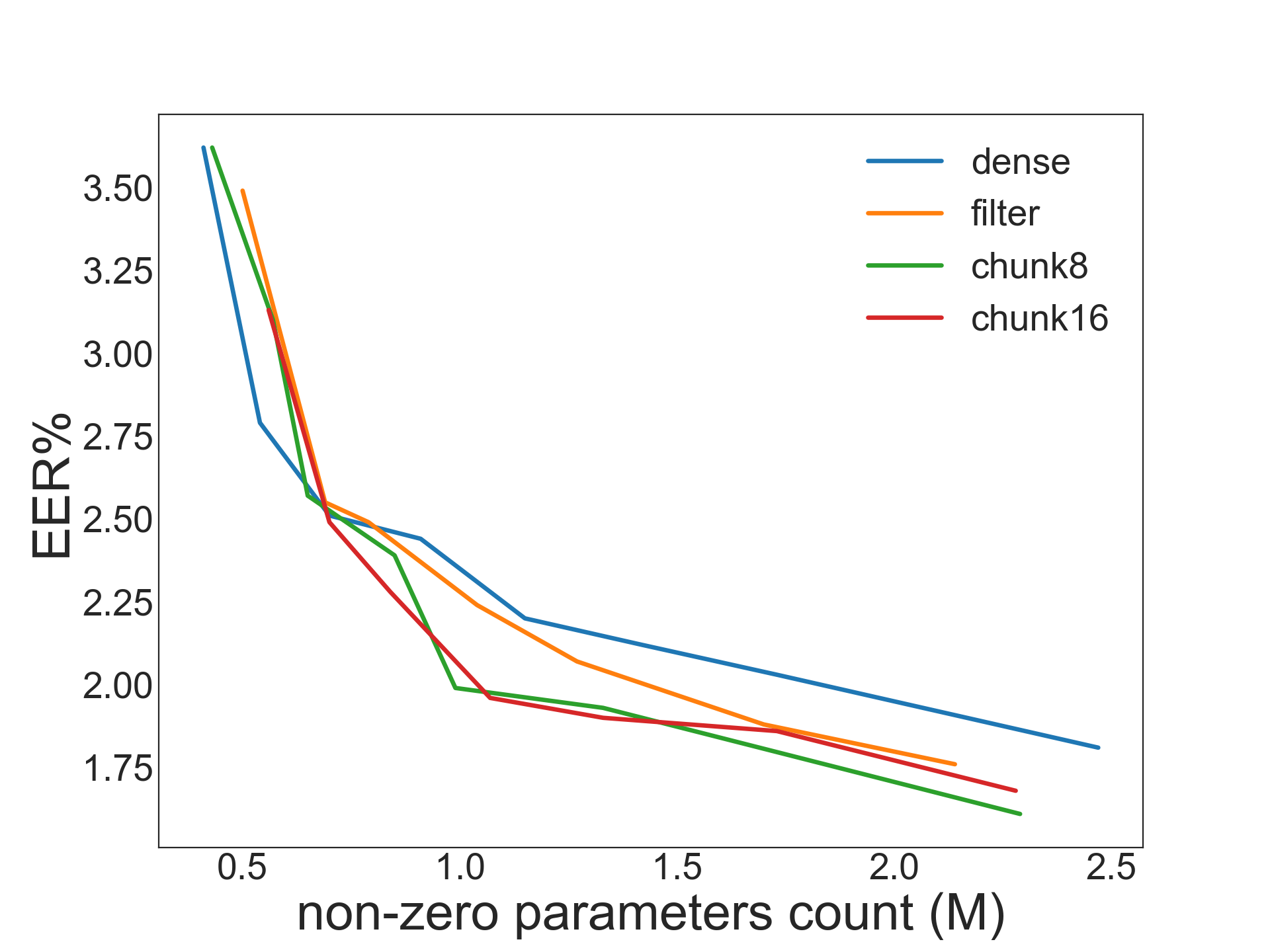}}
  \centerline{(a) {EER}}
\end{minipage}
\hfill
\begin{minipage}[b]{0.45\linewidth}
  \centering
  \centerline{\includegraphics[width=4.3cm]{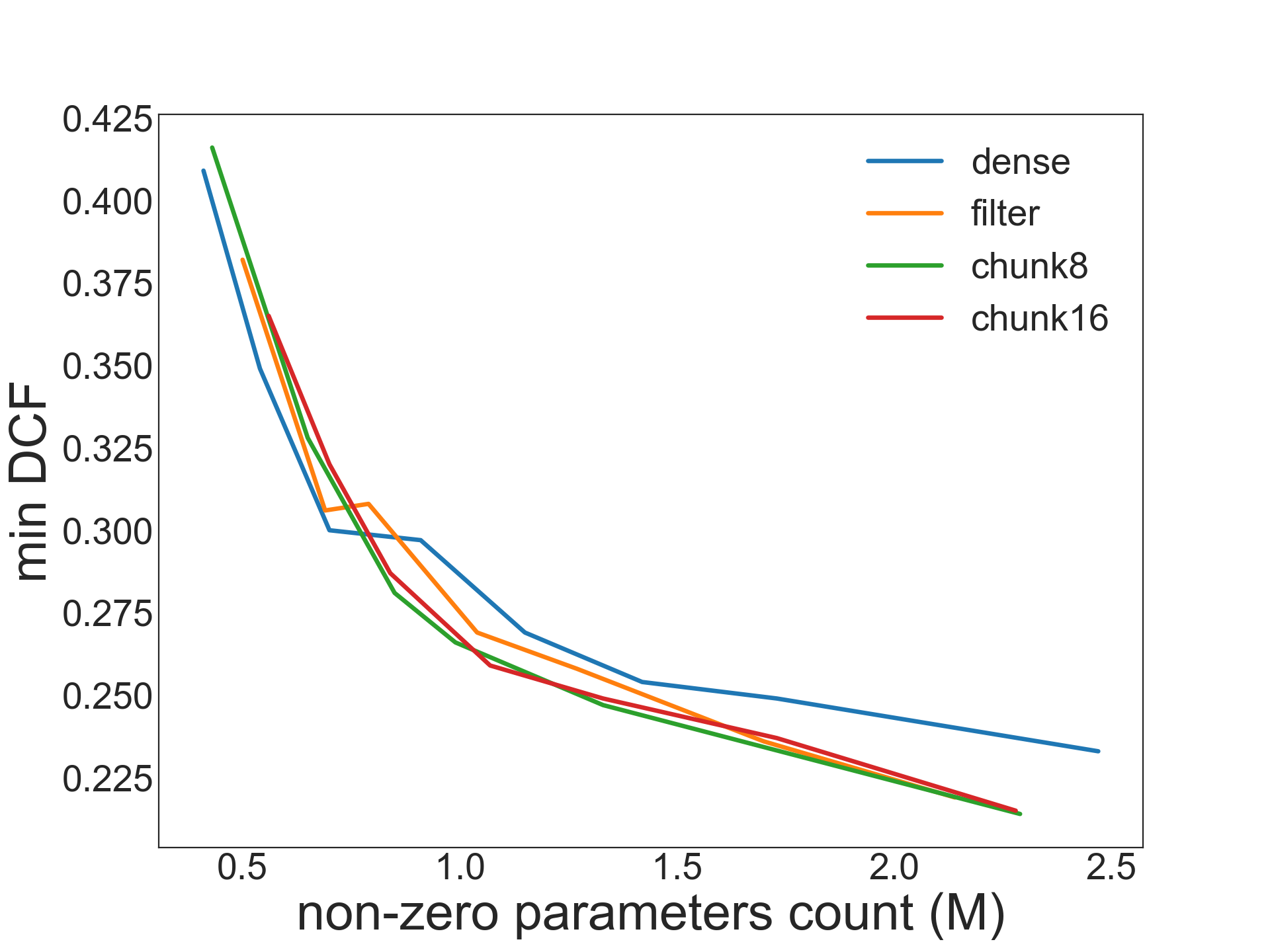}}
  \centerline{(b) {minDCF}}
\end{minipage}
\caption{Visualization of EER and minDCF versus parameter count in millions.}
\label{fig:lambda_EER}
\end{figure}

Figure~\ref{fig:lambda_EER} is the visualization of the relationship of non-zero parameters and EER or min detection cost function (minDCF) at $P_{target}=0.01$ (consistent with the VOiCES evaluation protocol). The X-axis represents the number of non-zero parameters and Y-axis is the EER and minDCF. We compared the filter sparsity, chunk-8 sparsity, and chunk-16 sparsity with dense models of different sizes. It is shown in Figure~\ref{fig:lambda_EER}(a) that when the number of parameters is large, sparse models achieve lower EER than dense models of the same size. However when the number of non-zero parameters is small, dense models have better performance. In our experimental setting, the turning point is around 0.7 million parameters. For example, with EER around 2.0\%, it is clear that models with smaller granularity have lower size. Chunk-8 models can reach 1.99\% EER with 0.99 million parameters and chunk-16 has 1.96\% EER under 1.07 million parameters. Comparing with baseline, smaller dense models reach 2.03\% EER with 1.73 million parameters, chunk-8 and chunk-16 both reach lower EER with less than 60\% of the parameters. Also, when non-zero parameter count is larger than 1.5 million, there is a tendency that chunk-8 has the best performance while filter sparsity has higher EER under the same non-zero parameter count. As for the relationship of minDCF, as is shown in Figure~\ref{fig:lambda_EER}(b), we observe similar patterns as seen in EER.  

A somewhat surprising finding in these results is that, filter\_1, chunk8\_1, and chunk16\_1 with less parameters have slightly lower EER, 1.76\%, 1.61\%, and 1.68\%, respectively, compared to the baseline of 1.81\%.  We believe this is because the group Lasso loss is an effective regularizer, and when used in a small dose, helps produce more generalized models.




\begin{table}[tb]
\caption{Model performance with different sparsity levels}
\label{tab:filters}
\small
\centering

\setlength{\tabcolsep}{1.5pt}
\vspace{2pt}
\begin{tabular}{ccccc|ccccc}
\toprule
Method & $\lambda$& size& EER& min& Method&$\lambda$ & size &EER&min\\
\textit{dense}&&(M)&(\%)&DCF&\textit{filter}&(e-2)&(M)&(\%)&DCF\\
\hline
baseline & - & 2.47 & 1.81 & 0.23 & baseline & - &2.47 & 1.81& 0.23 \\
\hline
dense\_1 & - & 1.73 & 2.03 & 0.25 & filter\_1 & 0.2  &2.14 & 1.76 & 0.22 \\
dense\_2 & - & 1.42 & 2.12 & 0.25 & filter\_2 & 0.5  &1.70 & 1.88 & 0.24 \\
dense\_3 & - & 1.15 & 2.20 & 0.27 & filter\_3 & 0.75 &1.27 & 2.07 & 0.26 \\
dense\_4 & - & 0.91 & 2.44 & 0.30 & filter\_4 & 1   &1.04 & 2.24 & 0.27 \\
dense\_5 & - & 0.70 & 2.51 & 0.30 & filter\_5 & 1.5  &0.79 & 2.49 & 0.31 \\
dense\_6 & - & 0.54 & 2.79 & 0.35 & filter\_6 & 2   &0.69 & 2.55 & 0.31 \\
dense\_7 & - & 0.41 & 3.62 & 0.41 & filter\_7 & 4   &0.50 & 3.49 & 0.38 \\
\bottomrule
\toprule
Method & $\lambda$& size& EER& min& Method&$\lambda$ & size &EER&min\\
\textit{chunk16}&(e-4)&(M)&(\%)&DCF&\textit{chunk8}&(e-4)&(M)&(\%)&DCF\\
\hline
baseline & - & 2.47 & 1.81 & 0.23 & baseline & - &2.47 & 1.81& 0.23 \\
\hline
chunk16\_1 & 0.25 & 2.28 & 1.68 & 0.22 & chunk8\_1 & 0.2   &2.29 & 1.61 & 0.21 \\
chunk16\_2 & 0.5  & 1.73 & 1.86 & 0.24 & chunk8\_2 & 0.5   &1.33 & 1.93 & 0.25 \\
chunk16\_3 & 1    & 1.33 & 1.90 & 0.25 & chunk8\_3 & 0.75 &0.99 & 1.99 & 0.27 \\
chunk16\_4 & 1.5  & 1.07 & 1.96 & 0.26 & chunk8\_4 & 1   &0.85 & 2.29 & 0.28 \\
chunk16\_5 & 2    & 0.84 & 2.28 & 0.29 & chunk8\_5 & 1.5 &0.65 & 2.57 & 0.33 \\
chunk16\_6 & 3    & 0.70 & 2.49 & 0.32 & chunk8\_6 & 2   &0.57 & 3.10 & 0.36 \\
chunk16\_7 & 4    & 0.56 & 3.13 & 0.37 & chunk8\_7 & 4   &0.43 & 3.62 & 0.42 \\
\bottomrule

\end{tabular}
\end{table}

\subsection{Measurements on Intel\textregistered\space GNA}
\label{ssec:simulate}

\begin{figure}
  \centering
  \centerline{\includegraphics[width=8cm]{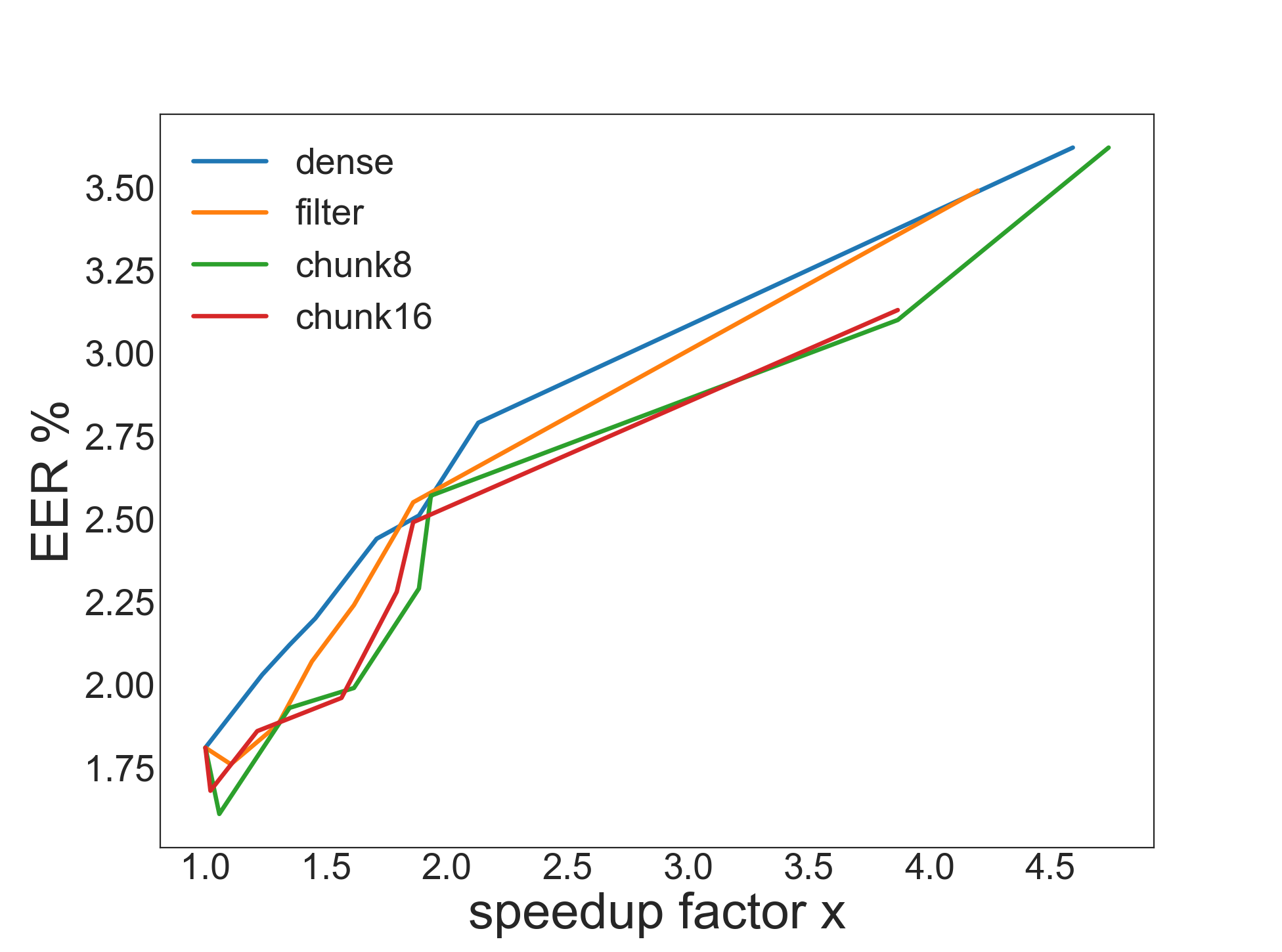}}
  \caption{Speedup on Intel\textregistered\space Pentium\textregistered\space Silver J4005}
  \label{fig:speedup}
\end{figure}

We also measured the actual inference time to find out how much speedup sparse models could achieve on Intel\textregistered\space GNA. We measured all the models we get in table~\ref{tab:filters} and found the relationship between the hardware speedup and EER. As is shown in Figure~\ref{fig:speedup}, four lines represent three different sparse granularity and one dense model and they all start at the baseline point. Generally, dense models always have higher EER when speedups are the same, which confirms our expectation. It means that under the same EER, structural sparse models are always faster than the dense model. It is also important to point out that when speedup is small, around $1.2\times$ speedup, sparse models have speedup even with lower EER, which is a free-meal. However, there may exist some oscillation when measuring the inference time in hardware so the results may not be precise. This may explain why sparse models show no benefit when speedup is around $1.9\times$ and the trend is consistent.

\section{Conclusion}
\label{sec:conclusion}
In this paper, we applied structural sparsification for speaker recognition models. By using pretrained models and group Lasso regularization, we kept the good performance of the original model while reducing the number of parameters and accelerating the actual execution.  For structural sparse models that are only slight smaller than the full size dense model, we achieved better performance on both EER and minDCF metrics.

\vspace{12pt}
\textbf{Acknowledgements.}
This work is supported by the National Science Foundation CCF-1910299.



\bibliographystyle{IEEEbib}
\bibliography{strings,refs}

\begin{thebibliography}{10}

\bibitem{heigold2016end}
Georg Heigold, Ignacio Moreno, Samy Bengio, and Noam Shazeer,
\newblock ``End-to-end text-dependent speaker verification,''
\newblock in {\em 2016 IEEE International Conference on Acoustics, Speech and
  Signal Processing (ICASSP)}. IEEE, 2016, pp. 5115--5119.

\bibitem{li2017deep}
Chao Li, Xiaokong Ma, Bing Jiang, Xiangang Li, Xuewei Zhang, Xiao Liu, Ying
  Cao, Ajay Kannan, and Zhenyao Zhu,
\newblock ``Deep speaker: an end-to-end neural speaker embedding system,''
\newblock {\em arXiv preprint arXiv:1705.02304}, 2017.

\bibitem{xvector}
David Snyder, Daniel Garcia-Romero, Gregory Sell, Daniel Povey, and Sanjeev
  Khudanpur,
\newblock ``X-vectors: Robust dnn embeddings for speaker recognition,''
\newblock in {\em 2018 IEEE International Conference on Acoustics, Speech and
  Signal Processing (ICASSP)}. IEEE, 2018, pp. 5329--5333.

\bibitem{richey2018voices}
Colleen Richey, Maria~A Barrios, Zeb Armstrong, Chris Bartels, Horacio Franco,
  Martin Graciarena, Aaron Lawson, Mahesh~Kumar Nandwana, Allen Stauffer,
  Julien van Hout, et~al.,
\newblock ``Voices obscured in complex environmental settings (voices)
  corpus,''
\newblock {\em arXiv preprint arXiv:1804.05053}, 2018.

\bibitem{nandwana2018robust}
Mahesh~Kumar Nandwana, Julien van Hout, Mitchell McLaren, Allen~R Stauffer,
  Colleen Richey, Aaron Lawson, and Martin Graciarena,
\newblock ``Robust speaker recognition from distant speech under real
  reverberant environments using speaker embeddings.,''
\newblock in {\em Interspeech}, 2018, pp. 1106--1110.

\bibitem{gna}
G.~Stemmer, M.~Georges, J.~Hofer, P.~Rozen, J.~Bauer, J.~Nowicki, T.~Bocklet,
  H.~R. Colett, O.~Falik, M.~Deisher, and S.~J. Downing,
\newblock ``Speech recognition and understanding on hardware-accelerated
  {DSP},''
\newblock in {\em Proc. Interspeech}, 2017, pp. 2036--2037.

\bibitem{ioffe2006probabilistic}
Sergey Ioffe,
\newblock ``Probabilistic linear discriminant analysis,''
\newblock in {\em European Conference on Computer Vision}. Springer, 2006, pp.
  531--542.

\bibitem{Han}
S.~Han, J.~Pool, J.~Tran, and W.~J. Dally,
\newblock ``Learning both weights and connections for efficient neural
  networks,''
\newblock in {\em Advances in Neural Information Processing Systems}, 2015.

\bibitem{Liu}
B.~Liu, M.~Wang, H.~Foroosh, M.~Tappen, and M.~Pensky,
\newblock ``Sparse convolutional neural networks,''
\newblock in {\em The IEEE Conference on Computer Vision and Pattern
  Recognition}, 2015.

\bibitem{zhang2019learning}
Jingchi Zhang, Wei Wen, Michael Deisher, Hsin-Pai Cheng, Hai Li, and Yiran
  Chen,
\newblock ``Learning efficient sparse structures in speech recognition,''
\newblock in {\em ICASSP 2019-2019 IEEE International Conference on Acoustics,
  Speech and Signal Processing (ICASSP)}. IEEE, 2019, pp. 2717--2721.

\bibitem{Wei_cnn}
W.~Wen, C.~Wu, Y.~W, Y.~Chen, and H.~Li,
\newblock ``Learning structured sparsity in deep neural networks,''
\newblock in {\em Advances in Neural Information Processing Systems}, 2016.

\bibitem{yuan2006model}
Ming Yuan and Yi~Lin,
\newblock ``Model selection and estimation in regression with grouped
  variables,''
\newblock {\em Journal of the Royal Statistical Society: Series B (Statistical
  Methodology)}, vol. 68, no. 1, pp. 49--67, 2006.

\bibitem{han2015deep}
Song Han, Huizi Mao, and William~J Dally,
\newblock ``Deep compression: Compressing deep neural networks with pruning,
  trained quantization and huffman coding,''
\newblock {\em arXiv preprint arXiv:1510.00149}, 2015.

\bibitem{Narang}
S.~Narang, G.~Diamos, S.~Sengupta, and E.~Elsen,
\newblock ``Exploring sparsity in recurrent neural networks,''
\newblock {\em arXiv:1704.05119}, 2017.

\bibitem{intrinsic}
W.~Wen, Y.~He, S.~Rajbhandari, W.~Wang, F.~Liu, B.~Hu, Y.~Chen, and H.~Li,
\newblock ``Learning intrinsic sparse structures within long short-term
  memory,''
\newblock {\em arXiv:1709.05027}, 2017.

\bibitem{Huang2019IntelFFSID}
Jonathan Huang and Tobias Bocklet,
\newblock ``{Intel Far-Field Speaker Recognition System for VOiCES Challenge
  2019},''
\newblock in {\em Proc. Interspeech 2019}, 2019, pp. 2473--2477.

\bibitem{wang2018additive}
Feng Wang, Jian Cheng, Weiyang Liu, and Haijun Liu,
\newblock ``Additive margin softmax for face verification,''
\newblock {\em IEEE Signal Processing Letters}, vol. 25, no. 7, pp. 926--930,
  2018.

\bibitem{openvino}
``Openvino toolkit,'' \url{https://docs.openvinotoolkit.org/},
\newblock Accessed: 2019-10-14.

\bibitem{nagrani2017voxceleb}
Arsha Nagrani, Joon~Son Chung, and Andrew Zisserman,
\newblock ``Voxceleb: a large-scale speaker identification dataset,''
\newblock {\em arXiv preprint arXiv:1706.08612}, 2017.

\bibitem{chung2018voxceleb2}
Joon~Son Chung, Arsha Nagrani, and Andrew Zisserman,
\newblock ``Voxceleb2: Deep speaker recognition,''
\newblock {\em arXiv preprint arXiv:1806.05622}, 2018.

\bibitem{scheibler2018pyroomacoustics}
Robin Scheibler, Eric Bezzam, and Ivan Dokmani{\'c},
\newblock ``Pyroomacoustics: A python package for audio room simulation and
  array processing algorithms,''
\newblock in {\em 2018 IEEE International Conference on Acoustics, Speech and
  Signal Processing (ICASSP)}. IEEE, 2018, pp. 351--355.

\bibitem{Snyder15-MAM}
David Snyder, Guoguo Chen, and Daniel Povey,
\newblock ``{MUSAN}: {A} {M}usic, {S}peech, and {N}oise {C}orpus,'' 2015,
\newblock arXiv:1510.08484v1.

\bibitem{gemmeke2017audio}
Jort~F Gemmeke, Daniel~PW Ellis, Dylan Freedman, Aren Jansen, Wade Lawrence,
  R~Channing Moore, Manoj Plakal, and Marvin Ritter,
\newblock ``Audio set: An ontology and human-labeled dataset for audio
  events,''
\newblock in {\em 2017 IEEE International Conference on Acoustics, Speech and
  Signal Processing (ICASSP)}. IEEE, 2017, pp. 776--780.

\bibitem{GNA2}
M.~Deisher and A.~Polonski,
\newblock ``Implementation of efficient, low power deep neural networks on
  next-generation intel client platforms,''
\newblock {\em http://sigport.org/1777}, 2017.

\end{thebibliography}

\end{document}